\documentclass[twocolumn]{aastex62}

\received{2019 April 18}
\revised{2019 May 9}
\accepted{2019 May 12}
\submitjournal{The Astrophysical Journal Letters}

\shorttitle{A circumplanetary disk around PDS~70~b?}
\shortauthors{Christiaens et al.}
\usepackage{amsmath}

\begin{document}

\title{Evidence for a circumplanetary disk around protoplanet PDS~70~b} 

\correspondingauthor{Valentin Christiaens}
\email{valentin.christiaens@monash.edu}

\author[0000-0002-0101-8814]{Valentin Christiaens}
\affil{School of Physics and Astronomy, Monash University, VIC 3800, Australia}
\nocollaboration

\author[0000-0002-3968-3780]{Faustine Cantalloube}
\affiliation{MPIA, Heidelberg, Germany}
\nocollaboration

\author[0000-0002-0433-9840]{Simon Casassus}
\affiliation{Departamento de Astronomia, Universidad de Chile, Santiago, Chile}
\nocollaboration

\author[0000-0002-4716-4235]{Daniel J. Price}
\affil{School of Physics and Astronomy, Monash University, VIC 3800, Australia}
\nocollaboration

\author[0000-0002-4006-6237]{Olivier Absil}
\altaffiliation{F.R.S.-FNRS Research Associate}
\affiliation{STAR institute, Universit\'e de Li\`ege, Li\`ege, Belgium}
\nocollaboration

\author[0000-0001-5907-5179]{Christophe Pinte}
\affiliation{School of Physics and Astronomy, Monash University, VIC 3800, Australia}
\nocollaboration

\author[0000-0001-8627-0404]{Julien Girard}
\affiliation{Space Telescope Science Institute, Baltimore, USA}
\nocollaboration

\author[0000-0003-0363-1492]{Matias Montesinos}
%%\affiliation{Instituto de F\'isica y Astronom\'ia, Universidad de Valpara\'iso, Chile}
\affiliation{Instituto de F\'isica y Astronom\'ia, Universidad de Valpara\'iso, Chile}
%\affiliation{Chinese Academy of Sciences South America Center for Astronomy, National Astronomical Observatories, Beijing, China}
\affiliation{CASSACA, National Astronomical Observatories, Beijing, China}
\affiliation{N\'ucleo Milenio de Formaci\'on Planetaria, Chile}
\nocollaboration

%\altaffiliation{Creator of AASTeX v6.1}
%\collaboration{(LaTeX collaboration)}



\begin{abstract}
% 198/200 words
%The formation of giant planets is expected to involve a disk of circumplanetary material.
We present the first observational evidence for a circumplanetary disk around the protoplanet PDS~70~b, %in the form of excess emission at wavelengths longer than $\sim 2.3\mu$m, 
based on a new spectrum in the $K$ band acquired with VLT/SINFONI.
We tested three hypotheses to explain the spectrum: Atmospheric emission from the planet with either (1) a single value of extinction or (2) variable extinction, and (3) a combined atmospheric and circumplanetary disk model. 
%We considered the BT-SETTL and Madhusudhan et al. (2011) grids of synthetic atmospheric spectra.
Goodness-of-fit indicators favour the third option, suggesting circumplanetary material contributing excess thermal emission --- most prominent at $\lambda \gtrsim 2.3 \mu$m. Inferred accretion rates ($\sim 10^{-7.8}$--$10^{-7.3} M_J$ yr$^{-1}$) are compatible with observational constraints based on the H$\alpha$ and Br$\gamma$ lines. %Best-fit parameters for the planet suggest 
For the planet, we derive an effective temperature of 1500--1600 K, surface gravity $\log(g)\sim 4.0$, radius $\sim 1.6 R_J$, mass $\sim 10 M_J$, and possible thick clouds. Models with variable extinction lead to slightly worse fits. %values of our goodness-of-fit indicators. 
However, the amplitude ($\Delta A_V \gtrsim 3$mag) and timescale of variation ($\lesssim$~years) required for the extinction would also suggest circumplanetary material.
%Simultaneous observations with a large wavelength coverage and including gas accretion tracers (e.g.~H$\alpha$) would enable to disentangle the second and third hypothesis.
\end{abstract}




%%% abrreviated citations
\defcitealias{Madhusudhan2011}{M11}
\defcitealias{Muller2018}{M18}
%\defcitealias{Keppler2018}{K18}
\defcitealias{Christiaens2019a}{C19}

%%%%%%%%%%%%%%%%% BODY OF PAPER %%%%%%%%%%%%%%%%%%

\section{Introduction}
Discovery of the Galilean moons of Jupiter \citep{Galilei1610} led to the overthrow of Geocentric cosmology, and Galileo's subsequent denouncement, trial and house arrest. Their hypothesised origin is a circumplanetary disk (CPD) of gas and dust \citep[e.g.][]{Lunine1982,Lissauer1993}. Analytic work \citep{Pollack1979,Canup2002,%Mosqueira2003,
Papaloizou2005} and increasingly sophisticated numerical simulations \citep[e.g.][]{Lubow1999,%DAngelo2002,
Ayliffe2009,Gressel2013,Szulagyi2017} have consolidated this hypothesis. 

 Observational evidence has so far remained elusive. Searches using mm-continuum have produced only non-detections \citep[e.g.][]{Isella2014,Wu2017,Wolff2017}. High-contrast infrared (IR) observations have been suggested instead to capture the thermal excess from the CPD \citep[e.g.][]{Zhu2015b,Eisner2015,Montesinos2015}. The power of high-contrast IR spectroscopy to characterize young substellar companions has been demonstrated over the last few years \citep[e.g.][]{Allers2013,Bonnefoy2014,Delorme2017}. Observed spectra are fitted to either synthetic atmospheric models or observed template spectra which enable the estimation of effective temperature, surface gravity and radius of the planet, which can then be used to estimate its mass and age. Constraints on clouds or haze in the atmosphere can also be obtained (e.g.~\citealt{Madhusudhan2011}, hereafter \citetalias{Madhusudhan2011}; see \citet{Madhusudhan2019} for a recent review).

%Mechanisms through which planets form are still poorly understood.
%Despite decades-long efforts on the theoretical side \citep[e.g.][]{Pollack1996,Boss2001}, the lack of direct observational data of planets caught at birth have quenched the selection and refinement of a particular type of formation model.
%Both the population of mature exoplanets probed with indirect detection methods \citep[e.g.][]{Winn2015} and the lack of giant planets directly imaged at large separations \citep[e.g.][]{Chauvin2013,Vigan2018} point in favor of the core-accretion mechanism.
%However, the inferred physical properties (e.g.~effective temperature and dynamical mass) of the few directly imaged adolescent giant planets appear more compatible with a hot-start formation scenario \citep[][]{Bonnefoy2013,Marleau2014,Samland2017}. %, which has been historically associated to formation through gravitational instability **REF***.
 
%Another aspect of giant planet formation is whether it involves the presence of a circumplanetary disk (CPD) of gas and dust.
%This hypothesis was first proposed to account for the observed properties of the Galilean moons and other major satellites of gas giants in the Solar system \citep[e.g.][]{Lunine1982,Lissauer1993}.
%Analytical developments and hydrodynamical simulations have further consolidated that hypothesis ***REF
%Nevertheless, these long-sought CPDs have remained elusive to date.
%Delorme+17: very dusty
We present evidence for a circumplanetary disk around the recently discovered protoplanet PDS~70~b. PDS~70 is a young K7-type star %{\bf of the Upper Centaurus-Lupus subgroup (UCL) \citep{Pecaut2016}} 
 surrounded by a pre-transitional disk with a large annular gap \citep{Hashimoto2012, Dong2012}. 
The protoplanet was first detected in this gap using VLT/SPHERE \citep[][hereafter \citetalias{Muller2018}]{Keppler2018,Muller2018}.
Subsequent search for sub-mm emission from a CPD %around PDS~70~b 
using ALMA was inconclusive \citep{Keppler2019}.
Spectral characterisation of the planet was presented in \citetalias{Muller2018}, suggesting an effective temperature of $\sim$1000-1500 K, surface gravity $\lesssim 3.5$dex and mass $\leq 17 M_{\rm J}$. We present a new spectral characterization of PDS~70~b, including both the measurements presented in \citetalias{Muller2018} and a new spectrum in $K$ band obtained with VLT/SINFONI \citep[][hereafter \citetalias{Christiaens2019a}]{Christiaens2019a}. The spectrum shows excess emission at $\lambda \gtrsim 2.3 \mu$m inconsistent with naked model atmospheres of young planets, indicating the presence of circumplanetary material.

\section{Observations}\label{Obs+DataRed}

\citetalias{Muller2018} presented a $YJH$ spectrum and multi-epoch broadband photometric measurements, %in the  filters $H1$, $H2$, $K1$ and $K2$, 
acquired with SPHERE/IFS ($YJH$), SPHERE/IRDIS ($H1$/$H2$ and $K1$/$K2$), NICI ($L'$) and NACO ($L'$). %We have re-estimated the contrast of the protoplanet with respect to the star in these datasets using ANDROMEDA \citep[][Cantalloube et al. 2019 in prep.]{Cantalloube2015} and found good agreement with the values quoted in \citet{Keppler2018} and \citetalias{Muller2018} for all measurements but the NACO $L'$ data. The contrast found with ANDROMEDA is $(8.55 \pm 3.38) \times 10^{-17}$W m$^{-2}$  $\mu$m$^{-1}$ and for the NICI and NACO datasets, respectively. 
We gathered all measurements quoted in \citetalias{Muller2018}, but considered updated values for their $L'$-band flux estimates. \citetalias{Muller2018} fitted the SED of the star to estimate its $L'$ flux, without including possible excess disk emission. That value was then multiplied by the contrast of the companion to infer its flux. However, most of excess IR emission compared to the star likely arises from hot dust in the inner disk %\citep[][]{Dong2012,Keppler2018,Keppler2019} 
that would be unresolved from the star at $L'$ band, and should hence be included. By contrast, the contribution from resolved scattered light from the disk at $L'$ band is negligible \citep[see images in][]{Keppler2018}. Given the importance of the thermal IR flux to the hypothesis of a CPD around PDS~70~b, we re-estimated the $L'$ flux of the companion considering %we first estimated 
(1) the $L'$ flux of the star + unresolved inner disk by interpolating the photometric measurements in the W1 ($3.35 \mu$m) and W2 ($4.60 \mu$m) filters of WISE \citep{Wright2010}, and (2) the same values for the contrast of the companion as in \citetalias{Muller2018}. The new $L'$ estimates of the companion are: $(8.55 \pm 3.38) \times 10^{-17}$ W m$^{-2}$ $\mu$m$^{-1}$ and $(6.70 \pm 4.05) \times 10^{-17}$ W m$^{-2} \mu$m for the NICI and NACO data, respectively.

In a recent paper \citepalias{Christiaens2019a}, we inferred the contrast of the protoplanet with respect to the star as a function of wavelength, $c(\lambda)$, in the $K$ band at unprecedented spectral resolution ($R\sim100$ after spectral binning) using VLT/SINFONI. We employed two different methods for extracting $c(\lambda)$, both leading to flux estimates consistent with each other at all wavelengths.
Here, we consider only the spectrum inferred with ANDROMEDA \citep[][Cantalloube et al. 2019 in prep.]{Cantalloube2015} since it has smaller uncertainties and higher spectral resolution, and avoids the risk of contamination by extended (resolved) disk signals.

Despite the higher quality of the ANDROMEDA $c(\lambda)$, some spectral channels contain outliers. In order to minimize the risk of bias, we first removed spectral channels with a detection below 3$\sigma$ and lying in strong telluric lines, %($\lesssim 1$\% of channels in the $K$ band), 
then used a Savitzky-Golay filter of order 3 with a 81-channel window to smooth the  $c(\lambda)$ curve before binning it by a factor of 20 \citep{Savitzky1964}.
We obtained the final $K$-band spectrum of the protoplanet by multiplying $c(\lambda)$ with the calibrated spectrum of the star measured with the SpeX spectrograph \citep{Long2018}, after resampling the latter at the spectral resolution of the binned SINFONI spectral channels.

In total, our SED has 86 data points; 49 from \citetalias{Muller2018} and 37 obtained with SINFONI \citepalias{Christiaens2019a}. Measurements span 6 years, extending from 2012/03 to 2018/02. Flux estimates at overlapping wavelengths are all consistent with each other except one.
%The spectrum inferred with SINFONI (\emph{yellow error bars} in Figure~\ref{FinalSpectrum}) is consistent with the estimates obtained with SPHERE/IRDIS in the $K1$ and $K2$ filters for the 2016/05 epoch, but only with the $K1$ filter for the 2018/02 epoch \citep[][\citetalias{Muller2018}]{Keppler2018}. 
Namely, our SINFONI flux estimates (2014/05 epoch) are slightly higher than the 2018/02 epoch SPHERE measurement in the $K2$ filter (2.25$\mu$m).

\section{Spectral analysis}
\label{Kspectrum}

We first attempted to fit the observed SED with synthetic spectra modeling pure atmospheric emission (Section~\ref{PhotosphericModels}). 
We considered a single value of extinction for all epochs, treated as a free parameter in the spectral fit (referred to as \emph{Type I models} hereafter).
Given the discrepancy between the SPHERE $K2$ and SINFONI $K$-band measurements, we also considered a fit with variable amount of extinction for different epochs (\emph{Type II models}).
Finally, we also examined models consisting of combined emission from an atmosphere and a circumplanetary disk with a variable accretion rate (\emph{Type III models}; Section~\ref{CPDModels}). %This type of fit is motivated by the redder slope seen in our SINFONI $K$-band spectrum than any synthetic model shown in \citetalias{Muller2018}.
In order to minimise the number of free parameters, we considered only two possible values of extinction (for Type II models) and two accretion rates (for Type III models); one value for the SINFONI (2014/05), SPHERE (2016/05) and NICI (2012/03) epochs, and the other for all other epochs. This division was chosen because (1) the SPHERE IFS ($YJH$)+IRDIS ($K1$/$K2$) were obtained simultaneously on 2018/02, and the IRDIS $H1$/$H2$ points of 2015/05 are consistent with the IFS spectrum, and (2) the SINFONI 2014/05 data are brighter than the IRDIS $K2$ point of 2018/02, while the IRDIS $K2$ point of 2016/05 is in better agreement with the SINFONI data. For the NICI and NACO points, we arbitrarily assigned them to the first and second group, respectively.

For all model types, we minimised the following goodness-of-fit indicator: %, as in \citet{Olofsson2016}: 
\begin{align}
\label{Eq:GoodnessOfFit}
\chi^2 = \sum_i \omega_i \big[ \frac{F_{\rm obs}(\lambda_i) - F_{\rm model} (\lambda_i)}{\sigma_i} \big]^2
\end{align}
where $\sigma_i$ is the uncertainty in the flux measurement $F_{\rm obs}(\lambda_i)$ at wavelength $\lambda_i$, and weights $\omega_i$ are defined for photometric and spectroscopic observations following a similar strategy as in \citet{Ballering2013} and \citet{Olofsson2016}. Weights are proportional to the FWHM of the filters used (for broadband photometric measurements), or the spectral resolution (for SPHERE/IFS and SINFONI data). The sum of all weights is normalized to the total number of points.
We define a \emph{reduced} goodness-of-fit indicator $\chi_r^2$ as $\chi^2$ divided by the respective number of degrees of freedom for each type of model.

%*** INSERT TABLE HERE WITH EXPLORED PARAMETERS FOR EACH MODEL***
%\begin{table*}
%\begin{center}
%\caption{Parameters explored for the spectral fit of PDS~70~b}
%\label{tab:ExploredParams}
%\begin{tabular}{lcccccccccc}
%\hline
%Model & $T_{\rm eff}$ & $\log(g)$ & $R_{\rm b}$ & $f$ & $c$ & $k$ & $A_V$ & $A_{V,2}^{\dagger}$ & $M_{\rm b}\dot{M_{\rm b}}$ & $M_{\rm b}\dot{M_{b,2}}^{\ddagger}$ \\
% & [K] &  & [$R_{\rm J}$] & & & & [mag] & [mag] & [$M_{\rm J}^2$] & [$M_{\rm J}^2$] \\
%\hline
%2012-03-31 & NICI & $L'$ & - & 99.4 & 118 & 1, 2 \\
%2014-05-10 & SINFONI & $K$ & $\sim$100 & 99.8 & 116 & 3\\
%2015-05-03 & IRDIS & $H2$/$H3$ & - & 52.0 & 70 & 2\\
%2015-05-31 & IRDIS & $H2$/$H3$ & - & 40.8 & 70 & 2 \\
%2016-05-14 & IRDIS & $K1$/$K2$ & - & 16.9 & 22 & 2 \\
%2016-06-01 & NACO & $L'$ & - & 83.7 & 155 & 2\\
%2018-02-24 & IFS & $YJH$ & 54 & 95.7 & 150 & 4\\
%2018-02-24 & IRDIS & $K1$/$K2$ & - & 95.7 & 150 & 4\\
%\hline
%\end{tabular}
%\end{center}
%$^{\dagger}$Second value of extinction considered in the fit allowing for variable extinction between the SINFONI epoch and other epochs.\\
%$^{\ddagger}$Second value of mass accretion rate considered in the allowing for variable accretion rate between the SINFONI epoch and other epochs.\\
%\end{table*}

\subsection{Atmospheric models}\label{PhotosphericModels}

\begin{figure*}
	\centering
	\includegraphics[width=\textwidth]{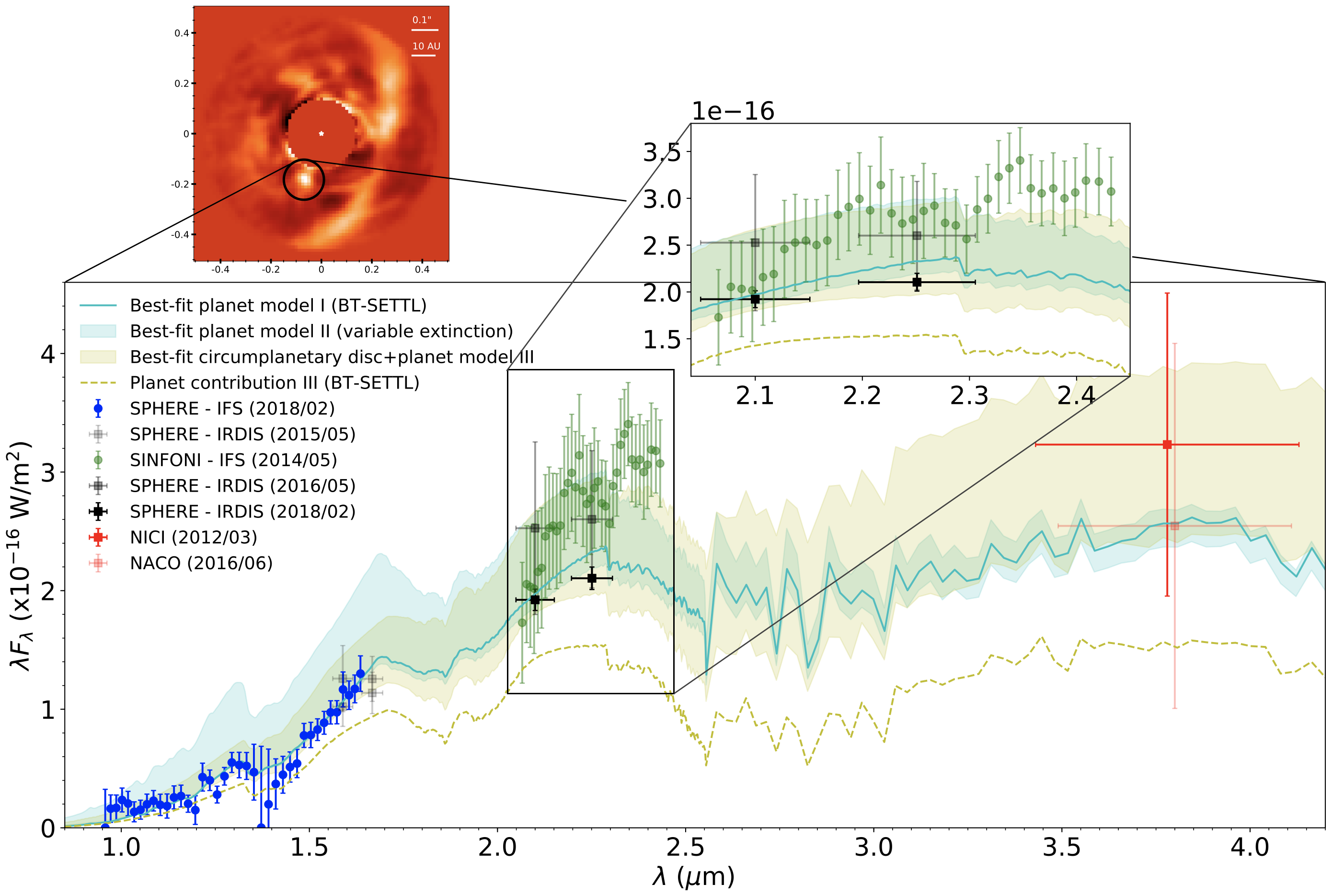}
    \caption{Combined spectrum of PDS 70 b compared to best-fit BT-SETTL models %, either consisting of photospheric emission alone ({\bf a}) or a combination of photospheric and circumplanetary disk emission ({\bf b}). 
    consisting of pure atmospheric emission without (solid cyan line) or with variable extinction (shaded blue area). Shaded yellow area shows best-fit atmosphere+CPD model. Circles with vertical error bars are IFS measurements, while squares with vertical and horizontal error bars are broad-band measurements. Inset highlights new SINFONI spectrum (green points). %See Table~\ref{tab:Observations} for details of each observation. %Compared to figure~6, the \emph{dotted lines} only consider points shortward of 1.7 $\mu$m for the fit. 
    %IFS data are given with \emph{circles}: \emph{yellow} for VLT/SINFONI - this work, and \emph{blue} for VLT/SPHERE \citep{Muller2018}. Broad-band photometric measurements are shown with \emph{squares}, where the horizontal bar represents the FWHM of the respective filter bandwidth. \emph{Light black} and \emph{black} squares correspond to the first and second epoch measurements with SPHERE/IRDIS, resp. \citep{Keppler2018,Muller2018}. \emph{Light red} and \emph{red} squares correspond to the $L'$ measurements with NACO and NICI, resp. \citep{Keppler2018}.
    %BT-SETTL models are taken from \citet{Allard2012,Baraffe2015}. SB12 stands for the best-fit model among all hot- and cold-start models presented in \citet{Spiegel2012}. The best-fit BT-SETTL and SB12 models are shown with \emph{black} and \emph{green} lines. \emph{Dashed lines} (BT-SETTL-1 and SB12-1) correspond to the best fit to all points of the spectrum, while \emph{dotted lines} (BT-SETTL-2 and SB12-2) only consider points shortward of 1.7 $\mu$m.
    %Best-fit BT-SETTL model \citep{Allard2012,Baraffe2015} is shown with a \emph{cyan} curve/shaded area, while best-fit model in the grid of synthetic atmospheric spectra presented in \citetalias{Madhusudhan2011} is given with a \emph{green} curve/shaded area. Search ranges and best-fit parameters are given in Table~\ref{tab:best_fit_params}.
    The SINFONI spectrum shows a $\sim 2 \sigma$ IR excess at $\lambda > 2.3 \mu$m compared to best-fit atmospheric models, best accounted for by the presence of a CPD. Top image shows image of PDS~70~b with SINFONI \citepalias[from][]{Christiaens2019a}.
    }
    \label{FinalSpectrum_BTSETTL}
\end{figure*}

\begin{figure*}
	\centering
	\includegraphics[width=\textwidth]{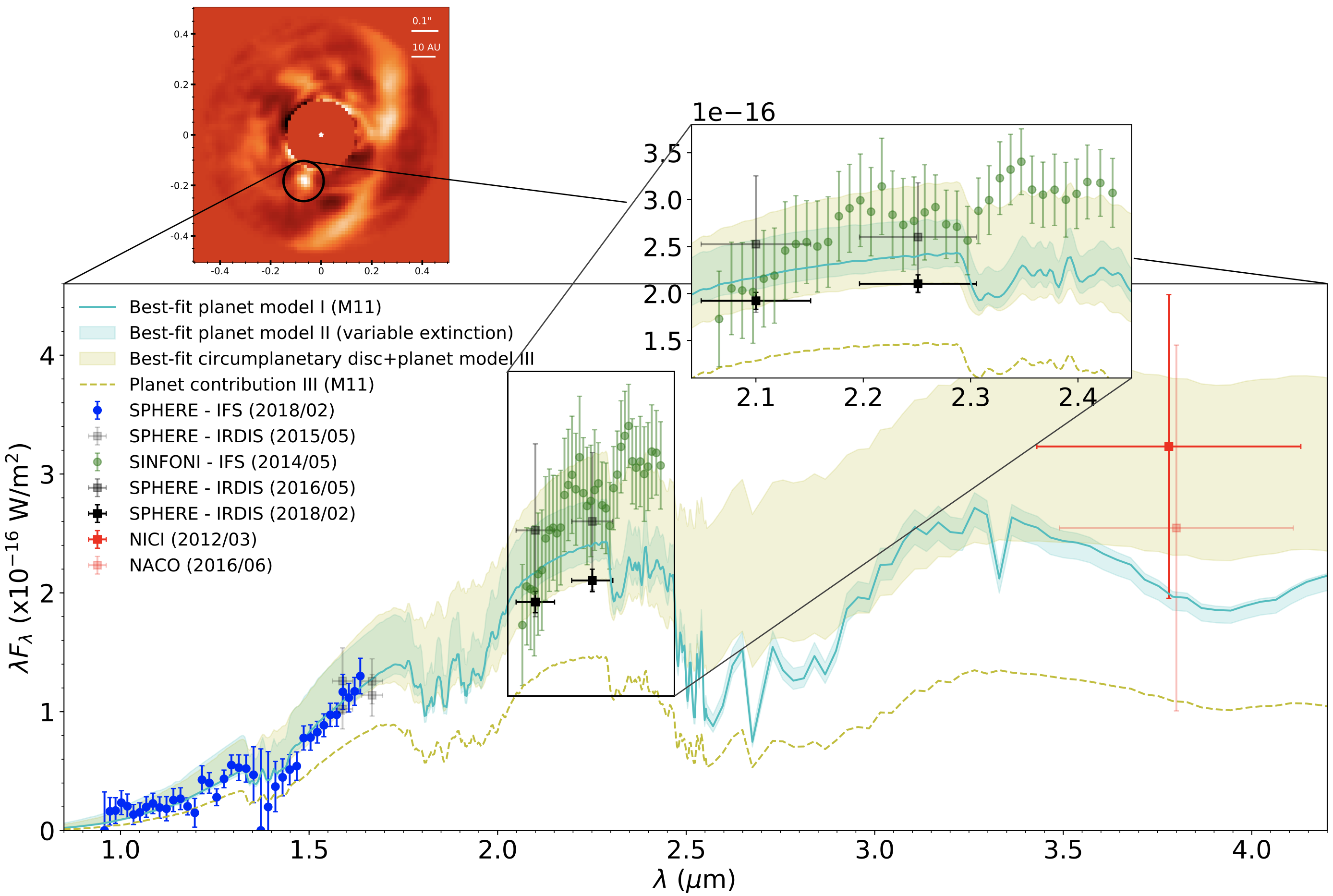}
    \caption{Same as Figure~\ref{FinalSpectrum_BTSETTL} but using \citetalias{Madhusudhan2011} models. Again, the best fit is obtained with a circumplanetary disc.
    %Spectrum of PDS 70 b (same symbols as in Figure~\ref{FinalSpectrum_BTSETTL}) compared to best-fit synthetic models consisting of a combination of atmospheric and circumplanetary disk emission. %Search ranges and best-fit parameters are given in Table~\ref{tab:best_fit_params}. 
    %Two values of accretion rates are allowed for (1) the SINFONI and SPHERE 2016/05 epochs and (2) the other epochs, respectively, but a single set of photospheric parameters is forced for all epochs. The models with the best-fit accretion rates correspond to the limits of the shaded \emph{cyan} (BTSETTL+CPD) and \emph{green} areas (\citetalias{Madhusudhan2011}+CPD). The atmospheric components are given with the \emph{dashed cyan} and \emph{dashed green} curves respectively. Assuming that part of the IR emission comes from an accreting CPD better accounts for all measurements than pure atmospheric models.
    }
    \label{FinalSpectrum_M11}
\end{figure*}

We considered two grids of synthetic spectra: BT-SETTL models \citep{Allard2012,%Allard2014,
Baraffe2015}, and the grid of atmospheric models presented in \citetalias{Madhusudhan2011}. % and the spectra associated to planet formation models presented in \citet[][hereafter SB12]{Spiegel2012}.
These models treat dust and clouds differently. BT-SETTL models account for dust formation using a parameter-free cloud model. They consider cloud microphysics --- the particle size of each species is calculated self-consistently based on condensation and sedimentation mixing timescales. Free parameters are the effective temperature, varied between 1200 and 1900 K (steps of 100 K), and the surface gravity, $\log(g)$, explored between 3.0 and 5.0 (steps of 0.5~dex). We assumed Solar metallicity.

\citetalias{Madhusudhan2011} considered a wide grid of cloud models, with different geometrical and optical thickness, particle size and metallicities. %SB12 considered a range of planet formation models between the coldest- and hottest-start models (set by the value of initial entropy), and computed associated spectra as a function of planet mass and age. %The initial entropy is a free parameter setting how hot or cold and is varied between 8.0 $k_B$/baryon and the minimum between 13.0 $k_B$/baryon and the maximum value for gravitational stability for each considered planet mass. 
%They consider both cloud-free and hybrid cloud models for either Solar or 3 dex larger than Solar metallicities. 
The \citetalias{Madhusudhan2011} models do not include microphysics. They consider different cloud spatial structure and particle sizes; labeled \emph{A}, \emph{AE}, \emph{AEE} or \emph{E}, based on the rapidity with which clouds are cut off at their upper end. %(see \citetalias{Madhusudhan2011}). 
Several modal particle sizes are considered, including 1, 60 and 100 $\mu$m. The grid also includes cloud-free models (\emph{NC}), with both equilibrium and non-equilibrium chemistry. In the latter case, two additional free parameters arise: the eddy diffusion coefficient $K_{zz}$ taking possible values of $10^2$, $10^4$ or $10^6$ cm$^2$ s$^{-1}$, and the sedimentation parameter $f_{\rm sed}$ \citep[defined as in][]{Ackerman2001}.
\citetalias{Madhusudhan2011} varied these parameters on a grid of effective temperature and surface gravity ranging from 600 to 1800 K (steps of 100 K) and from 3.5 to 5.0 (steps of 0.5 dex), respectively.

For both BT-SETTL and \citetalias{Madhusudhan2011} models, we treated the planet radius as a free parameter to scale the total emergent flux. We explored values between 0.1 and 5.0 $R_{\rm J}$ in steps of 0.1 $R_{\rm J}$. We also considered dust extinction as an additional free parameter, with allowed values between $A_V = 0$ and $10.0$ mag (steps of 0.2~mag). 
For Type II models, we explored both minimum and maximum extinction values within this grid.
We considered the extinction curve for interstellar dust \citep{Draine1989}.
Considering other dust species (e.g.~typical species found in the atmosphere of brown dwarfs) would increase the number of free parameters, and is not expected to give qualitatively different slopes after dereddening \citep[see e.g.][]{Marocco2014}.
%Furthermore, the visual identification of extended structures in the vicinity of the companion in high-contrast images of the system suggests that a large amount of dust might be present in the line-of-sight towards the protoplanet \citep{Muller2018,Christiaens2019a}. 
We assumed a distance of 113~pc \citep{Gaia2018}. 

%The best-fit M11 and BT-SETTL models reproduce overall better the observed spectrum than the best-fit SB12 model, probably for a better treatment of the clouds. 
%Nonetheless, the SB12 models are still interesting as they are directly related to a formation scenario.
%The best-fit SB12 model, with an inferred initial entropy of 10.25 $k_B$/baryon and mass of $3\pm1 M_{\rm Jup}$, appears most consistent with a hot-start formation scenario (see e.g.~figure 5 of SB12.
%The best-fit M11 and BT-SETTL effective temperatures (1100--1300 K) are in agreement with this conclusion.
%Similarly, the best-fit effective temperatures obtained with the M11 and BT-SETTL models (1100--1300 K) are 

% v2
\begin{table*}  
\begin{center}
\caption{Best fit parameters for PDS~70~b}%, for each type of spectral model.}
\label{tab:best_fit_params}
\begin{tabular}{lcclcclcc}
\hline
\hline
%\hline
\multicolumn{3}{c}{\bf I. Planet alone} & \multicolumn{3}{c}{\bf II. Planet alone (variable extinction)} & \multicolumn{3}{c}{\bf III. Planet and %variable
circumplanetary disc}\\
\hline
Parameter & Range & Best fit & Parameter & Range & Best fit & Parameter & Range & Best fit \\
\hline
\multicolumn{9}{c}{BT-SETTL atmospheric models}\\
\hline
$T_{\mathrm{eff}}$ [K] & 1200--1900 & 1500 & $T_{\mathrm{eff}}$ [K] & 1200--1900 & 1500 & $T_{\mathrm{eff}}$ [K] & 1200--1900 & 1500 \\
$\log(g)$ & 3.0--5.0 & 3.0 & $\log(g)$ & 3.0--5.0 & 3.0 & $\log(g)$ & 3.0--5.0 & 4.0\\ 
$R_{\rm b}$ [$R_{\rm J}$] & 0.1--5.0 & 2.2 & $R_{\rm b}$ [$R_{\rm J}$] & 0.1--5.0 & 2.1 & $R_{\rm b}$ [$R_{\rm J}$] & 0.1--2.0 & 1.6 \\
$A_V$ [mag] & 0.0--10.0 & 4.34 & $A_{V\mathrm{,1}}$ [mag] & 0.0--10.0 & 0.60 & $A_V$ [mag] & 0.0--10.0 & 6.40 \\
& & & $A_{V\mathrm{,2}}$ [mag] & 0.0--10.0 & 4.00 & $M_{\rm b}\dot{M}_{\rm b,1}$ [$M_{\rm J}^2$yr$^{-1}$] & $10^{-7}$--$10^{-6}$ & $10^{-6.4}$ \\
 & & &  & & & $M_{\rm b}\dot{M}_{\rm b,2}$ [$M_{\rm J}^2$yr$^{-1}$] & $10^{-7}$--$10^{-6}$ & $10^{-6.8}$\\
$M_{\rm b}^{\dagger}$ [$M_{\rm J}$] & -- & 1.9 & $M_{\rm b}^{\dagger}$ [$M_{\rm J}$] & -- & 1.7 & $M_{\rm b}^{\dagger}$ [$M_{\rm J}$] & -- & 9.9 \\
 & & & &  & & $\dot{M}_{\rm b}^{\dagger}$ [$M_{\rm J}$yr$^{-1}$] & -- & $10^{-7.8}$--$10^{-7.4}$ \\
%\multicolumn{3}{cc}{$\chi_r^2$ & 0.14} & \multicolumn{3}{cc}{$\chi_r^2$ & 0.14} & \multicolumn{3}{cc}{$\chi_r^2$ & 0.14} \\
%\multicolumn{3}{c}{$\chi_r^2$~~~~~~~0.14} & \multicolumn{3}{c}{$\chi_r^2$~~~~~~~0.14} & \multicolumn{3}{c}{$\chi_r^2$~~~~~~~0.14} \\
 & & $\chi_r^2 \sim 1.01$ & & & $\chi_r^2 \sim 0.52$ & & & $\chi_r^2 \sim 0.41$\\
\hline
\multicolumn{9}{c}{M11 atmospheric models}\\
%\hline
%Parameter & Range & Best fit value & Parameter & Range & Best fit value & Parameter & Range & Best fit value \\
\hline
$T_{\mathrm{eff}}$ [K] & 600--1800 & 1100 & $T_{\mathrm{eff}}$ [K] & 600--1800 & 1200 & $T_{\mathrm{eff}}$ [K] & 600--1800 & 1600  \\
$\log(g)$ & 3.5--5.5 & 4.0 & $\log(g)$ & 3.5--5.5 & 4.0 & $\log(g)$ & 3.5--5.5 & 4.0 \\ 
$R_{\rm b}$ [$R_{\rm J}$] & 0.1--5.0 & 3.3 & $R_{\rm b}$ [$R_{\rm J}$] & 0.1--5.0 & 2.8 & $R_{\rm b}$ [$R_{\rm J}$] & 0.1--2.0 & 1.6 \\
$c$ & [NC,E,A,...]$^{\ddagger}$ & A & $c$  & [NC,E,A,...]$^{\ddagger}$ & A & $c$ & [NC,E,A,...]$^{\ddagger}$ & A60$^{\ddagger}$ \\
$f_{\rm sed}$ & [eq.,0,1,2] & eq. & $f_{\rm sed}$ & [eq.,0,1,2] & eq. & $f_{\rm sed}$ & [eq.,0,1,2] & eq. \\
$K$ [cm$^2$s$^{-1}$] & [eq.,$10^2$--$10^6$] & eq. & $K$ [cm$^2$s$^{-1}$] & [eq.,$10^2$--$10^6$] & eq. & $K$ [cm$^2$s$^{-1}$]  & [eq.,$10^2$--$10^6$] & eq. \\
$A_V$ [mag] & 0.0--10.0& 3.00 & $A_{V\mathrm{,1}}$ [mag] & 0.0--10.0 & 0.80 & $A_V$ [mag] & 0.0--10.0 & 8.72 \\
& & & $A_{V\mathrm{,2}}$ [mag] & 0.0--1.2 & 4.00 & $M_{\rm b}\dot{M}_{\rm b,1}$ [$M_{\rm J}^2$yr$^{-1}$] & $10^{-7}$--$10^{-6}$ & $10^{-6.3}$ \\
 & & &  & & & $M_{\rm b}\dot{M}_{\rm b,2}$ [$M_{\rm J}^2$yr$^{-1}$] & $10^{-7}$--$10^{-6}$ & $10^{-6.7}$ \\
 $M_{\rm b}^{\dagger}$ [$M_{\rm J}$] & -- & 42.0 & $M_{\rm b}^{\dagger}$ [$M_{\rm J}$] & -- & 30.2 & $M_{\rm b}^{\dagger}$ [$M_{\rm J}$] & -- & 9.9 \\
 & & & &  & & $\dot{M}_{\rm b}^{\dagger}$ [$M_{\rm J}$yr$^{-1}$] & -- & $10^{-7.7}$--$10^{-7.3}$ \\
 & & $\chi_r^2 \sim 1.20$ & & & $\chi_r^2 \sim 0.70$ & & & $\chi_r^2 \sim 0.44$\\
%\multicolumn{3}{c}{$\chi_r^2$~~~~~~~0.14} & \multicolumn{3}{c}{$\chi_r^2$~~~~~~~0.14} & \multicolumn{3}{c}{$\chi_r^2$~~~~~~~0.14} \\
 \hline
\multicolumn{9}{l}{$^{\dagger}$ $M_{\rm b}$ is inferred from the best-fit $\log(g)$ and $R_{\rm b}$ values, and $\dot{M}_{\rm b}$ is inferred from $M_{\rm b}$ and the best-fit $M_{\rm b}\dot{M}_{\rm b}$.}\\
 \multicolumn{9}{l}{$^{\ddagger}$ See Section~\ref{PhotosphericModels} and \citetalias{Madhusudhan2011}. A60 refers to model cloud A (thickest) with a modal particle size of 60 $\mu$m. }\\
\end{tabular}
\end{center}
\end{table*}

Figures~\ref{FinalSpectrum_BTSETTL} and~\ref{FinalSpectrum_M11} compare our best-fit models from the BT-SETTL and \citetalias{Madhusudhan2011} grids, respectively, with the SED of PDS~70~b.
Table~\ref{tab:best_fit_params} gives the explored parameter ranges, best-fit parameter values and corresponding $\chi_r^2$, for each model. 
Best-fit Type I BT-SETTL and \citetalias{Madhusudhan2011} models reproduce most of the observed SED but are a poor match to the red end of the SINFONI spectrum, with a $>2\sigma$ discrepancy for most data points at wavelengths $\gtrsim 2.3 \mu$m. They lead to reduced goodness-of-fit indicators $\chi_r^2 \sim 1.01$ and 1.20. 

Allowing for variable extinction (Type II) yields best-fit models (shaded cyan areas in Figures~\ref{FinalSpectrum_BTSETTL} and~\ref{FinalSpectrum_M11}) in better agreement with flux estimates longward of $\sim 2.3 \mu$m ($\chi_r^2 \sim 0.52$ and 0.70). Upper and lower edges of the shaded areas correspond to the minimum and maximum extinction values ($A_{V,1}$ and $A_{V,2}$, in Table~\ref{tab:best_fit_params}) of our best-fit model, with $A_{V,1}$ better accounting for the 2014/05 SINFONI, 2016/05 SPHERE and 2012/03 NICI data points and $A_{V,2}$ accounting for data points at all other epochs. The difference in extinction is $A_V \gtrsim 3$ mag for both BT-SETTL and \citetalias{Madhusudhan2011} models.

For both Type I and II models, the best-fit effective temperatures (1100--1500 K), surface gravity ($\log(g)\sim3.0$--$4.0$), planet radius ($2.1$--$3.3R_J$) and hence mass (1.7--42.0 $M_J$) are in  approximate agreement with the previous estimates made in \citetalias{Muller2018}. Our mass estimates are uncertain because of the large steps in $\log(g)$ (0.5 dex) in our model grids, and hence do not rule out a brown dwarf. Our best-fit \citetalias{Madhusudhan2011} models correspond to the thickest cloud models (labelled \emph{A}); extending to the top of the atmosphere.
%Interestingly, all best-fit models appear to slightly underestimate the observed flux at the long end of the $K$ band ($\sim 2.3$ to $2.45 \mu$m).
%Comparison between best-fit models and observed spectrum suggests a slight excess between 
%This is better visualized in figure~\ref{FinalSpectrum_Hfit}, where we only considered the points shortward of 1.7$\mu$m for the fit.
%In addition to the best-fit models obtained considering all points of the spectrum (\emph{dashed lines} in figure \ref{FinalSpectrum}), we also considered the best fit obtained with only the points shortward of 1.7$\mu$m (figure~\ref{FinalSpectrum_Hfit}). This is because if the observed spectrum is indeed composed of both a photospheric and circumplanetary contribution, the latter is not expected to be significant shortward of the $K$ band \citep[e.g.][]{Zhu2015b,Montesinos2015}. The best-fit BT-SETTL model to points shortward of 1.7 $\mu$m (\emph{green dotted line} in figure \ref{FinalSpectrum}) suggests a more significant excess, which appears to be increasing with wavelength.

%Given the hints of possible excess thermal emission, we considered a joint-fit representing both the atmospheric and circumplanetary disk (CPD) emissions.

\subsection{Combined atmospheric+CPD models} \label{CPDModels}

For our Type III models (combined atmosphere+CPD emission), we considered the same two grids of atmospheric models as in Section~\ref{PhotosphericModels}, coupled with the CPD models presented in \citet{Eisner2015}. % for CPD emission: those presented in \citet[][hereafter Z15]{Zhu2015b} and \citet[][hereafter M15]{Montesinos2015}, respectively.
The latter add a single free parameter, the mass accretion rate, which sets the brightness and shape of the CPD spectrum. %and are similar to those presented in \citet{Zhu2015b}.
We explored values of mass accretion rates $\log(\dot{M}_b M_b [M_J^2 \textrm{yr}^{-1}])$ ranging from -7.0 to -6.0, in steps of 0.1~dex. We did not consider accretion rates smaller than $10^{-7}M_J^2$ yr$^{-1}$ because the corresponding models do not contribute significantly at NIR wavelengths. %, while accretion rates larger than $10^{-6}M_J^2$ yr$^{-1}$ lead to CPD models significantly brighter than the flux measured at $L'$ band ($3.8 \mu$m). 
We assume a fixed inner truncation radius of $2 R_{\rm J}$ in our CPD models, as in \citet{Eisner2015}. We thus truncated our grid of planetary radii to $2 R_{\rm J}$ for consistency. Other parameters were explored on the same grids as for Type I and II models.

Figures~\ref{FinalSpectrum_BTSETTL} and \ref{FinalSpectrum_M11} show the best-fit combined planet+CPD models (shaded yellow areas). Dashed lines show the contribution from the atmosphere alone. Upper and lower edges of the shaded areas correspond to maximum and minimum mass accretion rates of our best fit model ($M_b\dot{M}_{b,1}$ and $M_b\dot{M}_{b,2}$, respectively, in Table~\ref{tab:best_fit_params}), accounting for the 2014/05 SINFONI and 2016/05 SPHERE data points and, respectively, data points at all other epochs.
The planet+CPD best-fit models reproduce better the observed spectrum than pure atmospheric models (with or without variable accretion), with reduced goodness-of-fit indicators $\chi_r^2 \sim 0.41$ and 0.44 using BT-SETTL and \citetalias{Madhusudhan2011} models, respectively. 
Interestingly, the best-fit parameters for both the CPD and the planet are similar using either BT-SETTL or \citetalias{Madhusudhan2011} models: mass accretion rates ranging between $\sim 10^{-6.4}$ and $\sim 10^{-6.8}$ $M_J^2$ yr$^{-1}$, effective temperature of 1500--1600 K, surface gravity $\log(g)\sim 4.0$, radius of $\sim 1.6 R_J$, and mass of $\sim 10 M_J$. The \citetalias{Madhusudhan2011} best-fit model also suggests the thickest cloud geometry, with a modal particle size of 60 $\mu$m.

The estimate of 10 $M_J$ is larger than that inferred from the planet-only BT-SETTL models because the estimated $\log(g)$ is significantly larger, while the inferred $R_b$ is slightly smaller. For \citetalias{Madhusudhan2011} models the opposite is true because $\log(g)$ is similar but $R_b$ is smaller.
%{\bf The estimate of 10 $M_J$ is larger (resp.~smaller) than that inferred with the pure atmospheric BT-SETTL (resp.~\citetalias{Madhusudhan2011}) models because of a larger estimated $\log(g)$ and a slightly smaller $R_b$ (resp.~a constant estimated $\log(g)$ but smaller $R_b$)}.
%The uncertainty on the exact contribution from the CPD in the total observed flux suggest a larger uncertainty is to be associated with the planet mass estimated from $\log(g)$ and $R_b$ than for Type I and II models. However, w
This suggests an older planet when considering a CPD in the model. Interestingly, the estimated planet parameters ($T_{\rm eff}$, $\log(g)$, $R_b$, $M_b$) agree with the  BT-SETTL models for a mass of $10 M_J$ and age 9--11 Myr \citep[][]{Baraffe2015}\footnote{Available at \url{https://phoenix.ens-lyon.fr/Grids/BT-Settl/}}. %CIFIST2011_2015/ISOCHRONES}}. 
The inferred age is consistent with estimates for the star in %\citet[][%$\sim 8$Myr based on evolutionary models, and 9--16 Myr based on membership to UCL]{Pecaut2016},
\citet{Pecaut2016}, but not with the newer estimate of $5.4 \pm 1.0$ Myr \citepalias{Muller2018}. In contrast, parameters in planet-only models are inconsistent with \citet{Baraffe2015} evolutionary models for any combination of mass and age.

%Although similar to the models presented in \citet{Zhu2015b}, the \citet{Eisner2015} models 
%The main difference is that a single value of inner radius hence depend on the mass accretion rate of the planet (more specifically on $M_p \dot{M_p}$), the inner and outer radii of the CPD.
%The main difference is that the M15 models also consider the feedback of the planet luminosity on the disc. In their work. The M15 models are thus expected to be more appropriate to represent luminous protoplanets ($L \gtrsim 10^{-4} L_{\odot}$). The Z15 models are still expected to represent highly accreting but relatively faint planets (e.g.~cold start planets or low-mass hot-start giant planets).
%In order to reduce the number of free parameters, we fix the value of the outer radius of the CPD for both models, as it is not expected to influence significantly the SED of the CPD.

\section{Discussion} \label{Discussion}

We explored the hypothesis of variability for PDS~70~b because of the absence of atmospheric models red enough to account for both the 2018/02 SPHERE $K2$ measurement and the points at $\gtrsim 2.3 \mu$m in our 2014/05 SINFONI spectrum. This is further supported by the disagreement (albeit slight) between the SPHERE and SINFONI measurements at $\sim 2.25 \mu$m. 
Since the variability of classical T-Tauri stars is thought to be related to either variable amounts of extinction from intervening circumstellar dust \citep[e.g.][]{Bozhinova2016} %, Scholz2019} 
or irregular accretion \citep[e.g.][]{%Fernandez1996,
Bouvier2004,Rigon2017}, we tested similar hypotheses in our type II and III models, respectively. 
%Given the sampling of the SINFONI and SPHERE measurements, the possible variability of PDS~70~b 
%The time scale of variability for magnetospheric accretion on T-Tauri stars can be of the order of hours to months \citep[][and references therein]{Bouvier2004}. %see also Alencar et al. 2001, Bouvier et al. 2003
Accretion variability has also been predicted in magnetohydrodynamics simulations of forming planets \citep[e.g.][]{Gressel2013}, further justifying type III models.
%In case of magnetospheric accretion onto the protoplanet, an excess at shorter wavelengths ($\lesssim 1.5 \mu$m) corresponding to the heated photosphere is also expected to be observed \citep{Zhu2015b}. This does not appear to be observed in the SED of PDS~70~b, although significant extinction might complicate the detection of this excess. Some of our best-fit models yield estimates of extinction $A_V \gtrsim 3$ mag ***.

The best fit to the SED of PDS~70~b is obtained with an atmosphere+CPD model. Nonetheless, our caveats are: \begin{enumerate}
\item \emph{We considered a limited range of atmospheric models}. 
Atmospheric models have been proposed in recent years with different levels of complexity, including for example microphysics, non-equilibrium chemistry, or clouds/hazes %\citep[e.g.~PHOENIX, ExoREM, petit-CODE,][] 
\citep[see][and references therein]{Madhusudhan2019}.
We used the two most complete publicly available synthetic atmospheric model grids, which successfully reproduce the spectrum of adolescent giant exoplanets such as Beta Pic~b or HR~8799~b, c, d and e \citep[\citetalias{Madhusudhan2011};][]{Bonnefoy2013,Currie2014}.
Given that our best-fit type I (purely atmospheric) BT-SETTL and \citetalias{Madhusudhan2011} models are similar to the reddest atmospheric models (around $\sim 2.3 \mu$m) presented in \citetalias{Muller2018}, we do not expect our conclusion of an excess at $\gtrsim 2.3 \mu$m to change using different grids of atmospheric models.
%***Low-gravity can account for reddening effects, due to increased cloud optical depth \citep[e.g.][]{Marley2012, Charnay2018}.
%***NOt considered: different extinction laws corresponding to species different than interstellar dust. \citet{Marocco2014} considered the extinction laws of corundrum, forsterite and iron applied to the NIR spectrum of an L-type brown dwarf of similar effective temperature as PDS~70~b. Their results suggest that the IR excess observed at $\lambda \gtrsim 2.3 \mu$m cannot be accounted for by dust reddening alone.

%**Sec 4 of Zhu2015: Assuming magnetospheric accretion, LH$\alpha$ is proportional to $R_{\rm in}$. Small value inferred from Wagner+2018, although could be as high as 10-7 if Av ~ 3mag. Our non-detection of Br$\gamma$ in \citetalias{C19} suggests Av $<$ 3 mag. In that case $R_in$ is constrained to ***

\item \emph{The fit is not perfect}. 
%The most intriguing feature of the spectrum of PDS~70~b is the rise at $\sim 2.3 \mu$m. 
Although the best-fit atmosphere+CPD model best reproduces the excess at the end of the $K$ band, it does not perfectly reproduce the observed slope around 2.3$\mu$m. For both the BT-SETTL and \citetalias{Madhusudhan2011} type III models, most photometric points at wavelengths shorter than 2.3 $\mu$m lie below the model (albeit all within 2$\sigma$), while most points longward of 2.3 $\mu$m are slightly brighter than the model (but within 2$\sigma$ also).
%In addition, most atmospheric models with an effective temperature of 1100--1500 K predict a sharp drop at 2.29 $\mu$m corresponding to the first overtone of the CO bandhead \citep[a common feature in M and L dwarfs; e.g.][]{Cushing2005}.
%By contrast, the sharp rise observed between $\sim 2.3 \mu$m and $\sim 2.35 \mu$m may suggest the CO bandhead is seen in emission rather than in absorption (the first three CO transitions {\bf $v=2-0$, $v=3-1$ and $v=4-2$} are at 2.294, 2.323 and 2.352 $\mu$m, resp.). %For young accreting stellar objects, CO bandhead emission is expected (instead of absorption) for high luminosities and low mass accretion rates \citep{Calvet1991}. Alternatively, \citet{Martin1997} suggested that hot funnels of magnetospheric accretion onto T-Tauri stars could also create CO bandhead emission. Could any of these predictions hold for forming giant planets? CPD models in \citet{Ayliffe2009} and \citet{Szulagyi2017a} do predict high temperatures (up to several thousand K) in the inner regions, which might produce CO bandhead emission.

\item \emph{The assumptions behind our CPD models may be incorrect}.
Incorrect assumptions may explain why our CPD models do not reproduce perfectly the observed steep slope.
We fixed the inner truncation radius to 2$R_J$. As shown in \citet{Zhu2015b}, different inner truncation radii can lead to different predicted CPD spectra for a given mass accretion rate.
Furthermore, neither the models in \citet{Zhu2015b} nor in \citet{Eisner2015} take into account radiative feedback from the protoplanet itself. The best-fit effective temperature and radius found for PDS~70~b suggest a protoplanet luminosity $L_p\sim10^{-4} L_{\odot}$. %are of the order of $T_{\rm eff}\sim$1300 K and $R_{\rm b} \sim 2 R_{\rm J}$. 
\citet{Montesinos2015} showed that the effect of such bright protoplanet would be to further increase the IR excess of the CPD with a steeper spectral slope, hence possibly improving the match with the K-band spectrum of PDS~70~b. 
New dedicated simulations are required to verify this hypothesis.
%*** Here we did not consider CPD with radiative transfer feedback from the planet \citep[e.g.][]{Montesinos2015}. This is motivated by the large effective temperature of the companion. In comparison, considering the models presented in \citet{Zhu2015b} or \citet{Eisner2015}, which do not consider radiative feedback of the planet on the circumplanetary disk, would require a combination of accretion rate and inner truncation radius of CPD \citep[see e.g.~Figure 1 in][]{Zhu2015b} that appear incompatible with constraints on the accretion rate \citep{Wagner2018,Christiaens2019a} and radius of the companion (Table ***)
\end{enumerate}

%***The CPD models in \citet{Eisner2015} neglect the radiative feedback from the protoplanet on the CPD. luminosity assumed to be dominated by planet $L_p\sim10^{-4} L_{\odot}$ assuming an effective temperature of 1300 K and a radius of 2 $R_{\rm J}$, vs $L_{\rm acc} \approx 3\times 10^{-6} L_{\odot}$ assuming no extinction ().
%The contribution from accretion luminosity could be higher in case of significant extinction. For example, AV larger than *** would lead to similar contributions from the planet and the accretion luminosities.
%Considering this additional luminosity source might lead to an even brighter CPD at long IR wavelengths \citep[e.g.][]{Montesinos2015}.
However, %in addition to the best-fit model to the observed SED being the atmosphere+CPD model, 
several lines of evidence support the hypothesis of circumplanetary material:
\begin{enumerate}
\item \emph{Our best-fit accretion rates agree with observations.} %of independent gas accretion tracers.} 
Assuming a similar relationship between H$\alpha$ luminosity and mass accretion rate as T-Tauri stars, \citet{Wagner2018a} estimated the PDS~70~b accretion rate as $\sim 10^{-7.8} M_{\rm J}^2 R_{\rm J}^{-1}$ yr$^{-1}$ in the absence of extinction. However, the protoplanet is likely to be embedded within the circumplanetary material from which it feeds. %, as also suggested by the IR excess at $\gtrsim 2.3\mu$m in our spectrum. 
The observed accretion rate would then be $\sim 10^{-6.7} M_{\rm J}^2 R_{\rm J}^{-1}$ yr$^{-1}$ for $A_V \approx 3$mag. 
Assuming a similar relationship for Br$\gamma$ emission, \citetalias{Christiaens2019a} also constrained the accretion rate to be $< 10^{-6.2} M_{\rm J}^2 R_{\rm J}^{-1}$ yr$^{-1}$, considering negligible extinction at $K$ band.
Our best-fit models suggest strong extinction towards the protoplanet ($A_V > 6$mag), consistent with the presence of a CPD. Our estimates of mass ($M_b \sim 10 M_J$), accretion rates ($\sim 10^{-7.5} M_J$ yr$^{-1}$), extinction ($\gtrsim 6$ mag) and radii $R_{\rm b}$ ($\sim 1.6 R_J$), appear roughly compatible with both observational constraints. 
%Based on Figure~2 in \citet{Zhu2015b}, if accretion is magnetospheric and the disk is truncated by the magnetic field of the planet, the estimated mass accretion rate of $\sim 10^{-7.5} M_J$ yr$^{-1}$ and $R_{\rm in}/R_{\rm b}$ ratio (where $R_{\rm in}$ is the inner truncation radius of the CPD) would suggest that the magnetic field of PDS~70~b is $\lesssim 10$G.
Monitoring of the H$\alpha$ luminosity would confirm the variability of the accretion rate. %A hint of explanation might be found in \citet{Calvet1991}. In this work, protoplanetary disks with large irradiation compared to their mass accretion rate are expected to show CO bandhead in emission instead of absorption.
%The CPD models we use do not take stellar irradiation into account.

%\item \emph{Variable extinction remains a possibility}, although it would involve very large variations in extinction, which appear unlikely (*** for BT-SETTL and M11).
\item \emph{The photometric variability is only observed at relatively long NIR wavelength ($\sim 2.2 \mu$m).} %Best-fit models with variable extinction would also suggest the presence of circumplanetary material.} 
The best-fit models involving variable extinction lead to only slightly worse fits to the data than the atmosphere+CPD best-fit models. However, for the former models the amplitude of the variability is larger at short than at long NIR wavelengths, which is not observed despite multiple epoch observations at wavelengths shorter than 1.7 $\mu$m. %This hypothesis thus cannot be ruled out.
Even if the variability was due to extinction, given the radial separation of the protoplanet ($\sim 20$au), both the amplitude ($\Delta A_V \gtrsim 3$ mag) and timescale of extinction variability (less than several years) suggests it would also be caused by circumplanetary dust.

\item \emph{Tentative excess with respect to the atmosphere+CPD models at 2.29--2.35 $\mu$m might suggest CO bandhead emission.} For young stellar objects, CO bandhead emission ($\Delta v=2$; first transitions at 2.294, 2.323 and 2.352 $\mu$m) is %expected to arise from collisions in the hot and dense inner part of the circumstellar disc, in the case of high stellar luminosities and low mass accretion rates 
an indicator of disk presence \citep{Geballe1987,Davis2011}. %Alternatively, \citet{Martin1997} suggested that hot funnels of magnetospheric accretion onto T-Tauri stars could also create CO bandhead emission. Could any of these predictions hold for forming giant planets? 
CPD models in \citet{Ayliffe2009} and \citet{Szulagyi2017a} predict temperatures up to several thousand K, %in the inner regions, 
which might also produce CO bandhead emission.

%\item {\bf \emph{The apparent CO bandhead emission suggests the presence of hot and dense circumplanetary gas.} For young accreting stellar objects, CO bandhead emission is expected to arise from collisions in the hot and dense inner part of the circumstellar disc, in the case of high stellar luminosities and low mass accretion rates \citep{Geballe1987,Calvet1991}. Alternatively, \citet{Martin1997} suggested that hot funnels of magnetospheric accretion onto T-Tauri stars could also create CO bandhead emission. Could any of these predictions hold for forming giant planets? CPD models in \citet{Ayliffe2009} and \citet{Szulagyi2017a} do predict high temperatures (up to several thousand K) in the inner regions, which might produce CO bandhead emission.}

\item \emph{Presence of a spiral arm}. Our conclusion regarding the presence of circumplanetary material around PDS~70~b is consistent with recent images obtained with VLT/SINFONI, suggesting the presence of an outer spiral arm likely feeding the CPD \citepalias{Christiaens2019a}.

%\item Finally, while this work was in the final stages of review, \citet{Szulagyi2019} provided new IR flux predictions for CPDs, based on hydro-dynamical simulations followed by radiative transfer, which are in qualitative agreement with the CPD contribution we inferred, in the case of a $\sim 10 M_J$ planet.
\end{enumerate}

%In summary, we tested three types of models and obtained the best fit to the SED of PDS~70~b with a coupled atmosphere+circumplanetary disk model.
\section{Conclusions}
%\begin{enumerate}
    %\item 
    The SED of PDS~70~b is best fit by models that include a circumplanetary disc. Atmospheric models alone are not able to account for the observed flux at wavelengths $\gtrsim~2.3~\mu$m.
    %\item The sharp rise at $\sim 2.29$--$2.35 \mu$m might suggest CO bandhead emission{\bf , which would be another indication for the presence of a disk around PDS~70~b}.
    %\item 
    We infer an accretion rate of $10^{-7.8}$--$10^{-7.3} M_J$ yr$^{-1}$ for a $\sim~10~M_J$ planet with significant extinction, consistent with prior observations.
%\end{enumerate}
Simultaneous follow-up observations of PDS~70~b with wide spectral coverage in NIR including the H$\alpha$ line should confirm the scenario of variable accretion through a circumplanetary disc. 
%\section{Summary} \label{Summary}

%***IF ENOUGH SPACE***

\section*{Acknowledgments}

%VC and SC acknowledge support from the Millennium Science 
%Initiative (Chilean Ministry of Economy) through grant RC130007. 
%VC and OA acknowledge funding from the European Research Council under the European Union's Seventh Framework Programme (ERC Grant Agreement No.~337569) and from the French Community of Belgium through an ARC grant for Concerted Research Action.
We acknowledge funding from the Australian Research Council via DP180104235, FT130100034 and FT170100040. 
VC thanks %Sebastian Perez, Miriam Keppler and Roy van Boekel for constructive discussions. VC also acknowledges %Zachary Long and Mike Sitko for sharing their SpeX spectrum of PDS~70 %, 
Andre M\"uller for sharing the SPHERE spectrum of PDS~70~b. %, 
%We used the SpeX Prism Library and Analysis Toolkit, maintained by Adam Burgasser.
MM acknowledges financial support from the Chinese Academy of Sciences (CAS) through CASSACA-CONICYT  Postdoctoral Fellowship (Chile).

\vspace{5mm}
\facilities{VLT(SINFONI,SPHERE), MASSIVE (\url{www.massive.org.au})}


\software{ANDROMEDA \citep{Cantalloube2015}, VIP \citep{GomezGonzalez2017}, SpeX (\url{http://www.browndwarfs.org/spexprism})}

\bibliographystyle{aasjournal}
\bibliography{PDS70} % if your bibtex file is called example.bib

\end{document}